\documentstyle[12pt]{article}
\begin{document}

\newtheorem{theorem}{Proposition}
\newtheorem{lemma}{Lemma}

\begin{center}
{\Large{\bf A time-extended Hamiltonian formalism}} \\ 
\vspace{1cm}
{\large{Hasan G\"{u}mral}\footnote{hasan@gursey.gov.tr}

\vspace{1cm}

Feza G\"{u}rsey Institute  \\
P.O. Box 6, 81220 \c{C}engelk\"oy-\.Istanbul, Turkey} \\ 
\vspace{1cm}
\today

\end{center}
\vspace{15mm}

\section*{Abstract}

A Poisson structure on the time-extended space $I \times M$
is shown to be appropriate for a Hamiltonian formalism in which
time is no more a privileged variable and no {\it a priori}
geometry is assumed on the space $M$ of motions.
Possible geometries induced on the spatial domain $M$ are investigated.
An abstract representation space for $sl(2,R)$ algebra
with a concrete physical realization by the Darboux-Halphen system
is considered for demonstration.
The Poisson bi-vector on $I \times M$ is shown to possess two
intrinsic infinitesimal automorphisms one of which is known
as the modular or curl vector field.
Anchored to these two, an infinite hierarchy of automorphisms
can be generated.
Implications on the symmetry
structure of Hamiltonian dynamical systems are discussed.
As a generalization of the isomorphism between contact
flows and their symplectifications,
the relation between Hamiltonian flows on $I \times M$ and
infinitesimal motions on $M$ preserving a geometric
structure therein is demonstrated for volume preserving
diffeomorphisms in connection with three-dimensional motion
of an incompressible fluid.

\newpage

{\bf (1)}
The idea of incorparating time into geometric constructions
for dynamical systems in order to get rid of its distinguished role
in parametrizing the solution curves as well as those offered by 
the canonical Hamiltonian formalism
was mainly analysed in covariant framework \cite{roman}-\cite{per}.
Most of these approaches to time-dependent dynamical systems
are based on particular choices of either a symplectic or a
contact structure on the spatial domain $M$ of the flow depending on its
dimension and make use of the powerful techniques of covariant
geometry.

Owing to the fact that the basic object of a Hamiltonian formulation,
namely, the Poisson bracket, is associated with a bi-vector, we
expect to find a complete and natural understanding of Hamiltonian
geometries of time-dependent dynamical systems in a contravariant
treatment. With this motivation, together with the idea that a
time-dependent system becomes autonomous in one higher dimensions,
we wish to present a Hamiltonian framework for dynamical systems
in which time is no more a privileged variable and no {\it a priori}
geometry is assumed on the spatial domain $M$.
We shall utilize an appropriate algebraic representation
of dynamical systems as first order differential equations
and contravariant geometry of
Poisson structures on the time-extended space $R \times M$.

We shall show that the Hamiltonian formalism on a time-extended
setting has the property that it allows us to generate recursively
its infinitesimal automorphisms. This has consequences on the
symmetry structure of its Hamiltonian vector fields.
Applications to three-dimensional fluid motion in Lagrangian
description will enable us to obtain explicit realizations
of the infinite dimensional left Lie algebra of the group
of volume preserving diffeomorphisms generating the particle
relabelling symmetry and the associated conservation laws.

Although, the present formalism is particularly suited for
non-autonomous systems, it also gives interesting results for
autonomous dynamical systems with time-dependent invariants. We shall
show that the algebraic and geometric structure on the phase
space of the Darboux-Halphen system are implicit in its
time-dependent symmetries which are eventually what dictate
us to use a time-extended formalism.

\vspace{5mm}
{\bf (2)}
A non-autonomous dynamical system is represented by
a time-dependent vector field which is defined to be the map
$v : I \times M  \to TM$ such that for $t \in I \subset R$ and $m \in M$
we have $v(t,m) \in T_{m}M$. 
The flow of $v$ on $M$ 
is a one-parameter family $g_{t}$ of 
diffeomorphisms given by the solution of system of differential
equations
\begin{equation}
 {dx \over d t} 
 = { dg_{t}(x_{0}) \over d t}
 = (v_t \circ g_{t})(x_{0})
 = v_t(x)=v(t,x) \;,\;\;\; g_0=id_M             \label{tindep}
\end{equation}
where components of $x$ are local coordinates around $m$ and
$v(t,x)$ is the representative of $v$ in these coordinates.
Eqs.(\ref{tindep}) are equivalent
to the autonomous differential equations
\begin{equation}
 {dt \over d \tau} = 1\;,\;\;\;\;
 {dx \over d \tau} = v(t,x)             \label{tdepe}
\end{equation}
on $I \times M$ associated with the suspension 
\begin{equation}
    V(t,m)= \partial_{t}+v(t,m)  \in I \times T_{m}M    \label{tvel}  
\end{equation}    
of $v$ at the point $(t,m)$.
The idea of going up one higher dimensions can be seen to be
associated with the algebraic representation of first order
ordinary differential equations as a subvariety of the
first jet bundle over $I \times M$, identified with $I \times TM$
and embedded into $T(I \times M)$ \cite{saun}.
Each embedding corresponds to a different parametrization
of trajectories of $v$. Throughout, we shall consider only
the affine parametrization (\ref{tdepe}) with time $t$
which is taken to be the canonical coordinate on the open
interval $I \subset R$.

\vspace{5mm}
{\bf (3)}
By a Hamiltonian structure on a smooth $n$-dimensional manifold $N$
we shall mean a bilinear mapping
\begin{equation}
  \{ \;,\; \} : C^{\infty}(N) \times C^{\infty}(N) \to C^{\infty}(N) 
\end{equation}
on the space of smooth functions satisfying the conditions of
skew-symmetry $\{ f,g \} =- \{ g,f \} $ and the Jacobi identity
\begin{equation}
  \{ f, \{ g,h \} \} + \{ h, \{ f,g \} \} + \{ g, \{ h,f \} \} =0
\label{fgh}         \end{equation}
for arbitrary $f,g,h \in C^{\infty}(N) $, together with an isomorphism
into the Lie algebra of vector fields on $N$.
Note that, we do not require the vector field $\{ f, \; \}$ to
be a derivation on $C^{\infty}(N)$. This will enable us to deal
with local bracket of functions defined, for example,
by contact or conformally symplectic structures.

If, on the other hand, $\{ \;,\; \}$ satisfies the Leibniz' rule,
then it is a Poisson bracket and is associated with a bi-vector
field $P : N \to \Lambda^2(TN) = TN \wedge TN$ on $N$.
For $N=I \times M$ we take
\begin{equation}
   P = B \wedge \partial_{t} + E   \label{extp}
\end{equation}
where $B$ and $E$ are time-dependent vector and bi-vector fields
on $M$, respectively. The Jacobi identity (\ref{fgh}) for
$\{ \;,\; \}$ is equivalent to the vanishing of the Schouten
bracket $[P,P]$ of $P$ with itself.

The Schouten bracket is defined on the space $\Lambda(TN)=
\bigoplus_{k=0}^{n} \Lambda^k(TN)$ of multi-vector fields
over $N$ by
\begin{equation}    
    \omega ([P,Q])=(-1)^{pq+q} i(P)di(Q)(\omega)
                 +(-1)^{p} i(Q)di(P)(\omega)                
                  - d \omega(P \wedge Q)     
\end{equation}
where $P \in \Lambda^p(TN)$, $Q \in \Lambda^q(TN)$,
$\omega \in \Lambda^{(p+q-1)}(T^*N)$ and
$i(P)( \cdot)$ is the interior product \cite{lic}.
The bracket satisfies the properties
\begin{equation}    
    [ P,Q ] = (-1)^{pq} [ P,Q ]            \label{sch1}  
\end{equation}  \begin{equation}
  (-1)^{pr} [ [ P,Q ] , R ] +(-1)^{qp} [ [ Q,R ] , P ]     
      + (-1)^{rq} [ [ R,P ] , Q ] = 0            \label{sch2}   
\end{equation}  \begin{equation}
 \; [ P,Q \wedge R ] =  [ P,Q ] \wedge R  + (-1)^{pq+q} Q \wedge [ P,R ] \;
 \label{sch3}       \end{equation} 
which are the (graded) skew-symmetry, the Jacobi identity and the
Leibniz' rule, respectively. This turns $( [\;,\;], \Lambda(TN))$
into a graded Lie algebra, called the Schouten algebra \cite{izu}.

The Jacobi identity for $P$ can now be computed in a
coordinate independent way
using properties (\ref{sch1})-(\ref{sch3}) of the Schouten bracket.
\begin{theorem} (\ref{extp}) is a Poisson bi-vector
field on $I \times M$ if and only if
\begin{equation}
   [ E,E ] = 2 B \wedge {\partial E \over \partial t} \;,\;\;\;\;
   [ E,B ] = B \wedge {\partial B \over \partial t} \;.  \label{jacobi}
\end{equation}
\end{theorem}

\vspace{5mm}
{\bf (4)}
The Hamiltonian form of the suspended vector field
$V$ on $I \times M$ is
\begin{equation}
  V= \partial_{t}+v=  P(dh) =
          B(h) \partial_{t}+E(dh)-h_{,t}B        \label{eqm}
\end{equation}
where $h$ is a time-dependent conserved function of $v$.
Assuming $(V,P,h)$ is a Hamiltonian system, we shall
investigate the properties of $P$ in connection with the given
system $v$.
\begin{theorem}
Assume $(V,P,h)$ is a Hamiltonian system on $I \times M$.
Define the time-dependent bi-vector field $Q = E + v \wedge B$
on $M$. Then

(i) $Q$ is a one-parameter family of Poisson bi-vectors on $M$.

(ii) $Q$ and $P$ are compatible on $I \times M$.

(iii) $h$ is a Casimir function of $Q$.          \label{qq}
\end{theorem}
The proof consists of showing $(i)\;[ Q,Q ]=0$, $(ii)\;[ P,Q ]=0$
by a straightforward algebra using Eqs.(\ref{sch1}-\ref{sch3})
and, $(iii)\;Q(dh)=0$ follows from Eqs.(\ref{eqm}).
As a corollary, we see that any Hamiltonian system $(V,P,h)$
on $I \times M$ can always be reduced to $(V,P-Q=B \wedge V,h)$
which is independent of $E$.
The next result is the well-known linearization of
the Jacobi identity by Hamiltonian vector fields.
\begin{theorem}
Given a time-dependent system associated with $v$ on $M$,
the Poisson bi-vector (\ref{extp}) defining the Hamiltonian
structure (\ref{eqm}) of its suspension (\ref{tvel}) satisfies
the infinitesimal invariance conditions
\begin{equation}
 {\partial B \over \partial t}+ [ v,B ] = 0   \;, \;\;\;
 {\partial E \over \partial t}+ [ v,E ] = 
         B \wedge {\partial v \over \partial t}   \label{eje}
\end{equation}
which are equivalent to (\ref{jacobi}).
\end{theorem}
This gives a characterization of the vector field $B$ as a
time-dependent infinitesimal symmetry of $v$. Moreover,
it follows from second of Eqs.(\ref{jacobi}) that
$B$ is an infinitesimal automorphism of $P$.

\vspace{5mm}
{\bf (5)}
To summarize, the construction of Hamiltonian structure of
a given system amounts to solving the linear system consisting
of Eqs.(\ref{eje}), (\ref{eqm}) and the conservation law for $h$.
As a whole, the contravariant
picture relies on finding a time-dependent symmetry $B$.
The bi-vector field $E$ which is the pull-back of $P$ to
$M$ is, on the other hand, the key object determining the
induced Hamiltonian geometry on the flow domain $M$.
For given $v$ with an infinitesimal
symmetry $B$, $E$ can be solved from linear equations. The
solution satisfies the non-linear equations (\ref{jacobi})
which, in turn, define all possible (if any) Hamiltonian structures
on $M$. For example, if $[E,E]=0$ one has a Poisson structure
on $M$ which, if non-degenerate, is dual to a symplectic
structure. Another possible solution that results in
an algebra of functions on $M$ is given by
\begin{equation}
   [ E,E ] = 2 B \wedge E  \;,\;\;\;\;
   [ E,B ] = 0                             \label{js}
\end{equation}
and is called a Jacobi structure. Eqs.(\ref{js}) are the conditions
for the local bracket
\begin{equation}
  \{ f,g \} = E(df \wedge dg) + fB(g)-gB(f)
\end{equation}
of functions on $M$ to satisfy the Jacobi identity (\ref{fgh}).
In covariant geometry this corresponds to either a contact or a
conformally symplectic structure \cite{LM}.

\vspace{5mm}
{\bf (7)}
We shall now prove a distinguished invariance property
of time-extended Hamiltonian formalism.
Associated to any bi-vector $P$ on $N$ with volume
$n$-form $\nu$, there corresponds a vector field
$p_{\nu} \equiv D_{\nu}(P)$ where $D_{\nu} \equiv
\nu^{-1} \circ d \circ \nu$ is the curl operator
on $\Lambda (TN)$ introduced in Ref.\cite{kir} and \cite{koszul}.
If $P$ is a Poisson bi-vector $p_{\nu}$ is an infinitesimal
automorphism of $P$, called the modular vector field \cite{wei97},
and it is moreover, a Hamiltonian vector field.
Following result is a consequence of the decomposition
$N=I \times M$ adapted for our investigation of dynamical systems.
\begin{theorem}
For the Poisson bi-vector (\ref{extp}) the brackets
\begin{equation}
  [...[[B,p_{\nu}],p_{\nu}],...,]        \label{rec}
\end{equation}
generate recursively an infinite hierarchy of automorphisms.  \label{rr}
\end{theorem}
{\bf Proof:} We consider the Jacobi identity
(\ref{sch2}) for the triple $(P,B,p_{\nu})$ of multi-vector fields
\begin{equation}
  0= [[P,B],p_{\nu}]+[[B,p_{\nu}],P]-[[p_{\nu},P],B]  \label{jrec}
\end{equation}
where the first and the last terms vanish because 
$B$ and $p_{\nu}$ are automorphisms. So, $[B,p_{\nu}]$ is
also an automorphism. Using this result in eq.(\ref{jrec})
with $B$ replaced by $[B,p_{\nu}]$, we arrive at the conclusion.
$\bullet$

\vspace{5mm}
{\bf (8)}
The explicit recursive construction of invariances of $P$
has immediate implications on the symmetry structure of
Hamiltonian vector fields associated with $P$. To this end,
we want to relate the automorphisms of $P$ to the infinitesimal
symmetries of Hamiltonian vector fields. Since
$P(dh)$ and $p_{\nu}$ are automorphisms, so is their bracket 
\begin{equation}
  [p_{\nu},P(dh)]=P(d(p_{\nu}(h)))   \label{sym1}
\end{equation}
which is Hamiltonian as well.
We observe that if $p_{\nu}(h)$ is a Casimir function of $P$,
then $p_{\nu}$ is a time-dependent infinitesimal symmetry of $v$.
Thus, in this case, the infinitesimal automorphisms of $P$ generated
recursively by (\ref{rec}) can be carried, together with
their Lie algebraic structure, over the algebra of
infinitesimal symmetries of $v$. By construction, the symmetry
algebra of $v$ may well be infinite dimensional. However,
the Casimir condition and hence the degeneracy of $P$
brings restrictions on these symmetries.
When $P$ is non-degerate so that it has only
trivial Casimirs, we can set $p_{\nu}(h)=0$.

The symmetry algebra consisting of vector fields on $I \times M$
of the form $U_k= \xi_k \partial_t + u_k$ can equivalently
be represented by time-dependent vector fields on $M$.
These are the unique characteristic forms
$\hat{u}_k=u_k-\xi_kv$ of $U_k$'s along the given
vector field $v$ \cite{olver}.
To generate them one replaces $p_{\nu}$ in Eq.(\ref{rec}) with
\begin{equation}
  \hat{p}_{\nu}= D_{\nu}(Q)+div_{\nu}(V)B   \label{ccurl}
\end{equation}
where $Q$ is the bi-vector introduced in proposition(\ref{qq}).
It follows that if the volume $n$-form $\nu$ is invariant
under the flow of $v$, the characteristic form of the curl
of Poisson bi-vector $P$ on $I \times M$ is the curl of the
Poisson bi-vector $Q$ on $M$ which is compatible with $P$.

\vspace{5mm}
{\bf (9)}
The geometric framework at our disposal finds applications
in qualitative analysis of geometric features of the flow space $M$.
We presented in \cite{hg96} an example due to Weinstein \cite{alan} 
in which $M$ has the structure of the dual space of Lie algebras.  
In Weinstein's example a Poisson structure on $I \times {\bf g}^{*}$
reduces to a parametric family of Lie-Poisson structures on duals of 
three dimensional Lie algebras having different topological structures
for certain values of the parameter.
Along the same line of inquiry, we shall now consider 
an example in which $M$ has the structure of a Lie algebra.

Let $M$ be the Lie algebra $sl(2,R)$ or, more generally, any smooth
manifold having an action of it. Let the vector fields $v,u,w$ satisfying
the Lie bracket relations
\begin{equation}
   [ v,u ] =-2v \;,\;\; [ w,u ] =2w \;,\;\; [ v,w ] = u
\label{sl2}      \end{equation}
be a basis of $sl(2,R)$ or fundamental vector fields of the
action on $M$. Then, $v \wedge u $ and $ w \wedge u$
define on $M$ two incompatible Poisson structures and the pair
$ E=w \wedge v ,\; B=u $ result in a Jacobi structure.

Defining the time-dependent vector fields $U_{t}=u+2 t v $ and
$W_{t}=-w+ t u+ t^{2}v $ on $M$, we find that $(-v,U_{t},W_{t})$
constitute a parametric family of basis for $sl(2,R)$.
Moreover, they satisfy the conditions
\begin{equation}
   [ \partial_t +v , U_{t} ]=0 \;,\;\;\;
   [ \partial_t +v , W_{t} ]=0
\end{equation}
to be infinitesimal symmetries of $v$. Including the generator
$\partial_t$ of one-dimensional algebra of translations into
the time-dependent basis, we obtain on $I \times M$ the
structure of a semi-direct sum of algebras.

On the time-extended space $I \times M$ the Poisson bi-vectors
$P_1= (\partial_{t} +v) \wedge U_{t}  $ and
$  P_2 = (\partial_{t} +v) \wedge W_{t} $ are compatible. 
$P_1$ reduces to $v \wedge u$ on $M$ whereas
the induced structure by $P_2$ is a one-parameter family
of Jacobi structures 
\begin{equation}
   E_{t} =  v \wedge (-w + t u)   \;,\;\;\;  B_{t}= u+2tv
\end{equation}
which includes $( w \wedge v ,\; u) $ for $t=0$.

\vspace{5mm}
{\bf (10)}
A concrete physical realization in coordinates of the above
abstract algebraic setting is provided by the Darboux-Halphen
system \cite{darb}
\begin{equation}
  v(m) = (yz-xy-xz)  \partial_x   + (xz-xy-yz) \partial_y  + 
         (xy-xz-yz) \partial_z    \label{hal}
\end{equation}
possessing the time-dependent symmetry transformations \cite{hal}
\begin{equation}
  (t,x^{i}) \mapsto ( {a t + b \over c t + d },
   2 c {c t + d \over a d - c b}
     +{(c t + d)^{2} \over a d - c b}x^{i})
\label{haltr}     \end{equation}
where $(x,y,z)$ is a local coordinate system around $m$
and $a,b,c,d \in R$ with $ad-cb \neq 0$. We refer to Ref.\cite{chak}
and the references therein for the recent resurrection of this
system in modern theoretical physics.

The lifts of the generators on $I \times M$ of the three-parameter
family of transformations (\ref{haltr}) along the vector $v(m)$
result precisely in the time-dependent basis $(-v(m),U_{t}(m),W_{t}(m))$
of $sl(2,R)$ at the point $m \in M$.
Here, $v(m)$ is given by (\ref{hal}) and we find that $u,w$
have the representatives
\begin{equation}
  u(m) =2(x \partial_x  + y \partial_y  +z \partial_z)  \;,\;\;\;
  w(m) = \partial_x  + \partial_y + \partial_z      \label{bos}
\end{equation}
in the adapted coordinate system \cite{gn93}.
Thus, starting from the transformations
(\ref{haltr}) and reading the above abstract algebraic construction
backward, we recover the geometric and algebraic structure
on flow space of the autonomous system (\ref{hal}).

\vspace{5mm}
{\bf (11)}
In covariant framework the construction of a symplectic structure
on (even dimensional) $I \times M$ relies on finding a necessarily
degenerate invariant two-form on $M$. In this case, the construction
of proposition(\ref{rr}) follows from 
\begin{theorem}
Let $\omega$ be a time-dependent, closed two-form on $M$. If it is
a relative invariant of $v$ then $\partial_t+v$ is symplectic. \label{covs}
\end{theorem}
{\bf Proof:} The closure and invariance conditions imply that
$d(i(v)(\omega)-\sigma)=0$ for some one-form $\sigma$. By Poincar\'e
lemma there exist a function $h$ such that $dh=i(v)(\omega)-\sigma$.
Solving its time dependence from $\varphi_{,t}=i(v)(\sigma)$ makes $h$
into a conserved quantity for $v$ and results in Hamiltonian
form of $\partial_t+v$ with the two-form $\sigma \wedge dt+\omega$.
This is closed on $I \times M$ provided the time dependence of
$\omega$ is determined from $\omega_{,t}+d\sigma=0$. $\bullet$

For a three-dimensional smooth manifold $M$ the covariant
and contravariant approaches to the construction of Hamiltonian
structure on $I \times M$ become identical. Namely, for
$M \equiv R^3$, $\omega$ and $\sigma$ 
can be associated with three component vector fields
$\omega ={\bf B} \cdot d{\bf x} \wedge d{\bf x}$ and
$\sigma = {\bf E} \cdot d{\bf x}$
where ${\bf B}$ is divergence-free because $\omega$ is closed.
Then, the equations in the proof of proposition(\ref{covs}) become
\begin{equation}
  \nabla \varphi = - {\bf v} \times {\bf B} - {\bf E} \;,\;\;\;\;
  \varphi_{,t}= {\bf v} \cdot {\bf E}  \;,\;\;\;
  {\bf B}_{,t}+ \nabla \times {\bf E} =0
\end{equation}
the last of which is equivalent to the symmetry condition on
${\bf B}$ when ${\bf E}$ is solved from the first.

\vspace{5mm}
{\bf (12)}
In hydrodynamical context \cite{arn},\cite{mr}, ${\bf B}$ can be
recognized to be
a so-called frozen-in field and the above construction enables
one to obtain the symplectic structure of suspended velocity field 
in Lagrangian description of fluid motion, that is, the description
by trajectories of $v$. Thus, in Lagrangian coordinates $\partial_t +v$
where $v$ is governed by a general force field,
is Hamiltonian with $\varphi$ and the symplectic two-form
\begin{equation}
  \Omega = {\bf B} \cdot d{\bf x} \wedge d{\bf x}
  - (\nabla \varphi + {\bf v} \times {\bf B})  \cdot d{\bf x} \wedge dt   
        \label{sympl}   
\end{equation}
for which $\rho_{\varphi} \equiv - {\bf B} \cdot \nabla \varphi$ is
the Liouville density of the invariant symplectic volume.
This formal symplectic set-up has been used to construct Lagrangian
and Eulerian invariants of hydrodynamic flows as well as to elucidate
the connection between them. The following results has been
obtained in \cite{hg97},\cite{hg98}.

The vector fields $U_k=({\cal L}_{U_1})^k(U_0)$ where
$U_0= \rho_{\varphi}^{-1} B$ and
$$  U_1= -U_0(h) (\partial_{t} +v) + f U_0 + W_1  \;,\;\;\;
 W_1 \equiv \rho^{-1}_{\varphi}  \nabla \varphi
   \times \nabla h  \cdot \nabla    $$
is the Hamiltonian vector field for the time-dependent function
$h$ satisfying $dh/dt=f(\varphi)$ for some function $f$,
are infinitesimal symmetries of $v$.

If $h$ is a conserved function of $v$, so are $- U_0(h_k)$
where $h_k \equiv  -(W_1)^{k-2}(U_0(h)) $.
In this case, the characteristic vector fields associated with
$U_k$'s reduce to the left-invariant vector fields
$W_k \equiv  ({\cal L}_{W_1})^{k-1}(U_0)$ which are
$\rho_{\varphi}$-divergence-free.
$U_k, W_k$ and $U_k - W_k$ are Hamiltonian vector fields with
the common Hamiltonian function $h_k$ and the Poisson bi-vectors
$P \equiv \Omega^{-1}, Q$ and $P-Q$, respectively.
The Poisson bracket algebra on $M$ of functions $h_k$ with the
bracket defined by $Q$ is isomorphic to the algebra of vector
fields $W_k$ generating the time-dependent volume preserving
diffeomorphisms (known as particle relabelling symmetries) on $M$.

\vspace{5mm}
{\bf (13)}
We shall now show that the formal symplectic structure (\ref{sympl})
can also be used to study the geometry of invariants of $v$.
In particular, we shall consider the Eulerian conservation law
of helicity and its degeneration into a Lagrangian invariant
in the geometric language of
Hamiltonian structures on $I \times M$. 
We assume that ${\bf B} = \nabla \times {\bf A}$
for some vector potential ${\bf A}$.
Then the symplectic two-form (\ref{sympl}) is exact
$\Omega = -d \theta$ with $- \theta =  \psi \, dt + {\bf A} \cdot d{\bf x}$.
The canonical one-form $\theta$ is a relative invariant of $v$
\begin{equation}
   {\cal L}_{\partial_{t}+u}( \theta )= d \chi \;,\;\;\;
   \chi \equiv  \psi + \varphi + {\bf v} \cdot {\bf A}
\end{equation}
and it becomes an absolute invariant if $\chi$ is a constant.
The three form $\theta \wedge d \theta$ is also a relative invariant 
\begin{equation}
   {\cal L}_{\partial_{t}+u}( \theta \wedge d \theta)=
           d (\chi \Omega )
\end{equation}
and it results in (magnetic) helicity conservation of Eulerian
description via the identity
$d(\theta \wedge d \theta)- \Omega \wedge \Omega =0$.
When $\chi =constant$, $\theta \wedge d \theta$ becomes an
absolute invariant and this turns helicity into a Lagrangian invariant,
that is, conserved function of the velocity field $v$.

Consider now the two-forms $\Omega_{\chi} \equiv \chi \Omega$
for $\chi \neq constant$.
Since $\Omega$ is closed they satisfy the conditions
\begin{equation}
   d \Omega_{\chi} = \eta  \wedge \Omega_{\chi}
   \;,\;\;\;\;\;  d \eta =0   \;,         \label{consym}
\end{equation}
$\eta \equiv d log \chi $, to define a 
conformally symplectic structure on $I \times M$ \cite{LM}.
Thus, in addition to the symplectic structure $\Omega$, we have
the family $\Omega_{\chi}$ of conformally symplectic
structures on $I \times M$ that coincide with $\Omega$
on the hypersurfaces $\chi =constant$ in $I \times M$
on which Eulerian conserved densities become Lagrangian invariants.

Denoting the bi-vector dual to $\Omega$ by $P$, one finds that
the contravariant version of (\ref{consym}) is the
Jacobi structure defined by the pair 
\begin{equation}
   P_{\chi} \equiv {1 \over \chi} P \;,\;\;\;\;
   W_{\chi} \equiv - {1 \over \chi^{2}} P(d \chi)   \label{fjacobi}
\end{equation}
and is conformally equivalent to $P$. We therefore conclude
that the absolute invariance of the canonical one-form,
the degeneration of Eulerian conservation law of helicity into
a Lagrangian one
and, the conformal equivalence of the local structure (\ref{consym})
to the symplectic structure (\ref{sympl}) as well as of their
contravariant versions are all the same.

\vspace{5mm}
{\bf (14)}
We have presented a conceptually straight and technically natural
approach to the Hamiltonian structure of 
dynamical systems without any assumption on time dependence
of either the systems themselves or their invariants and,
on dimensionality or geometry of the spatial domain.
We showed, with examples, various applications of the time-extended
Hamiltonian formalism.
Application to the Darboux-Halphen system indicates that 
this framework may be useful to obtain interesting results
even for autonomous systems.
We conclude with the following remarks on the time-extended
Hamiltonian formalism and its applications.

\begin{itemize}
\item
The time-extended framework provides alternative approaches
to problems arising from the distinguished role of time
either its use in parametrization of solution curves
or those offered by the canonical Hamiltonian formalism
as well as to geometric constructions involving time variable.
The construction is free from any particular parametrization
of solution curves on $M$ and is suitable for the study
of reparametrized systems.
\item
All possible Hamiltonian geometries on the flow
space $M$, including the local Lie algebra of functions,
can be obtained from the Poisson structure on $I \times M$.
This is in contrast to the approaches to time-dependent
systems that specify a contact or symplectic structure
on $M$ according to its dimensionality \cite{roman},\cite{per}.
Since we have $H_{DR}(R \times M) = H_{DR}(M)$ for the cohomology
spaces \cite{bott}, one can even think of exploiting the
structure on $I \times M$ to study the global geometric
properties of the flow space $M$.
\item
If the Hamiltonian structure on $I \times M$ is obtained
as a symplectification of a contact structure on $M$, one
can find an isomorphism between contact flows on $M$ and
Hamiltonian flows on $I \times M$ \cite{LM}.
The present framework generalizes this correspondence with
Hamiltonian flows on $I \times M$ to other infinitesimal
motions on $M$ preserving a given geometric structure.
The motion of an incompressible fluid is a manifestation of
this with the diffeomorphisms preserving the symplectic volume.
\item
The modular vector field is intrinsically associated to any
Poisson bi-vector on an oriented manifold. It is an element
of the algebra of infinitesimal Poisson automorphisms and
measures the extent to which Hamiltonian vector fields
are divergence free \cite{wei97}. A Poisson bi-vector
on $I \times M$ provides us with another intrinsic
infinitesimal automorphism and there follows the recursive
contruction of infinitely many of them.
\item
The idea is then to realize the symmetries of the motion
on $M$ in the automorphism group of the Poisson structure
on $I \times M$. This is similar to the way one obtains
infinitely many symmetries from a bi-Hamiltonian structure
\cite{magri},\cite{olver} or, invariants of geodesic
motions from the isometries of metric tensor \cite{mtw}.
In all cases one sets up a realization of symmetries
of infinitesimal motions (vector fields) in the invariance
group of a geometric structure (tensor field(s)).
Thus, the time-extended Hamiltonian formalism results
in another manifestation of this philosophy of
Felix Klein with the technology of Sophus Lie \cite{gam}.

\end{itemize}

\end{document}